\documentclass[12pt]{iopart}

\usepackage{iopams}

\usepackage{graphicx,fixltx2e}
\usepackage{lineno}
\usepackage{textcomp}  

\begin{document}

\title[3D FSLE and coherent structures in turbulent flows]{Characterization of coherent structures in three-dimensional
turbulent flows using the finite-size Lyapunov exponent}

\author{Jo\~{a}o H Bettencourt, Crist\'obal L\'{o}pez
and Emilio Hern\'{a}ndez-Garc\'{\i}a}

\address{IFISC (CSIC-UIB), Instituto de F\'{\i}sica Interdisciplinar y Sistemas Complejos  \\
              Campus Universitat de les Illes Balears\\
              E-07122 Palma de Mallorca, Spain}
\ead{joaob@ifisc.uib-csic.es}
\begin{abstract}

In this paper we use the finite size Lyapunov Exponent (FSLE) to 
characterize Lagrangian coherent structures in three-dimensional 
(3d) turbulent flows. Lagrangian coherent structures act as the organizers of
transport in fluid flows and are crucial to understand their
stirring and mixing properties. Generalized maxima (ridges) of the FSLE fields 
are used to locate these coherent structures.

Three-dimensional FSLE fields are calculated in two phenomenologically 
distinct turbulent flows: a wall-bounded flow (channel 
flow) and a regional oceanic flow obtained by numerical solution of 
the primitive equations where two-dimensional turbulence 
dominates. 

In the channel flow, autocorrelations of the FSLE field show 
that the structure
is substantially different from the near wall to the mid-channel 
region and relates well to the more widely studied Eulerian 
coherent structure of the turbulent channel flow. The ridges
of the FSLE field have complex shapes due to the 3d
character of the turbulent fluctuations. 

In the oceanic flow, strong horizontal stirring is present and the flow regime
is similar to that of 2d turbulence where the domain is populated
by coherent eddies that interact strongly. This in turn results in the 
presence of high FSLE lines throughout the domain leading to strong non-local 
mixing. The ridges of the FSLE field are 
quasi-vertical surfaces, indicating that the horizontal dynamics dominates
the flow. Indeed, due to rotation and stratification, vertical motions
in the ocean are much less intense than horizontal ones. This suppression is 
absent in the channel flow, as the 3d character of the FSLE ridges shows.
\end{abstract}

\submitto{\JPA}
\maketitle

\linenumbers

\setlength\linenumbersep{7pt}

\section{Introduction}

Turbulent flow occurs in the natural environmental
and in technological applications with such frequency that 
it could be considered the "natural" state of fluid flows 
to be found around us.
Traditionally, fluid flows have been observed and studied in 
the Eulerian perspective where a fixed
position is observed for a definite interval of time. The other 
perspective, the Lagrangian, follows the motion of the 
fluid and thus is better suited to study aspects of fluid flow
such as material transport or the deformation of fluid material
in a given state of motion.  

The use of stretching quantifiers such as the Lyapunov
exponents, which measure the relative separation between
particles
\cite{Artale1997,Aurell1997,Boffetta2001,Lapeyre2002}, has
broadly improved the Lagrangian study of fluid flows. 
On the one hand Lyapunov methods provide information on time scales
for dispersion processes, with its relevance for mixing and
stirring of fluids
\cite{Artale1997,Aurell1997,Boffetta2001,Boffetta2000,dOvidio2004,Poje2010}.
On the other, they are useful to detect the so-called
Lagrangian coherent structures (LCS). LCSs \cite{Haller2000,
Haller2000b} are templates for particle advection in complex
flows, separating regions with different dynamical behavior and
acting as barriers and avenues to transport,
fronts or eddy boundaries
\cite{Haller2000b,Boffetta2001,Lapeyre2002,Joseph2002,dOvidio2004,Mancho2006b,dOvidio2009,Peacock2010}.

Relationships of LCSs with Lyapunov fields have been 
established for the case of finite-time
Lyapunov exponents (FTLEs) \cite{Shadden2005,Lekien2007}. 
These relationships state that LCS can be identified 
with the ridges (generalized maxima) of the FTLE field.
Furthermore they state that the flux through the LCS is inversely 
proportional to the strength of the ridge and to the integration time
of the FTLE field calculation. This flux is shown to be small and the 
LCS extracted as the ridges of FTLE fields are considered to be almost
material-like surfaces. This identification has become widely used in the 
field although it should be mentioned that there are other more precise
definitions of LCS \cite{Mancho2006b,Haller2011,Haller2012},
that consider LCS to be exact material surfaces admitting zero flux 
across them. 
In our work, we use instead
finite-size Lyapunov exponents (FSLEs), which quantify the
separation rate of fluid particles between two given distance
thresholds \cite{Artale1997,Aurell1997}. They turn out to be
convenient for the case of bounded flows in which
characteristic spatial scales are more direct to identify than
temporal ones and have been shown to be robust with respect
to noisy or poorly resolved velocity fields \cite{Ismael2011}. 
Although a rigorous connection between the FSLE
and LCSs has not been established yet, previous work \cite{Joseph2002,dOvidio2004,
Molcard2006,dOvidio2009,Branicki2009} has shown that the ridges 
of the FSLE behave in a similar fashion as the ridges of the 
FTLE field. Following these works we assume that LCSs can be
computed as ridges of FSLEs, and that they are transported by
the flow as almost material surfaces/lines, with negligible flux 
of particles through them. Observations presented here are 
consistent with those assumptions.

Despite its relevance in real flows, the full three-dimensional
(3d) structure of LCSs is still an open subject. In 3d flows,
LCS were explored in atmospheric contexts \cite{DuToit2010,
Tang2011,Tallapragada2011}, and in a turbulent channel flow at
$Re_{\tau}=180$ in \cite{Green2007}. A kinematic ABC flow was
studied in \cite{Haller2001}. In the ocean, where it is widely
recognized that filamental structures, eddies, and in general
oceanic meso- and submeso-scale structures have a great
influence on marine ecosystems
\cite{Bakun1996,Garccon2001,Levy2001,Levy2008}, the
identification of LCSs and the study of their role in the
transport of biogeochemical tracers has primarily been
restricted to two-dimensional (2d) layers  
\cite{Rossi2008,Rossi2009,TewKai2009,Olascoaga2010}. There
are two concurrent reasons for this: a) because of
stratification and rotation, vertical motions in the ocean are
usually very small when compared to horizontal displacements;
b) synoptic measurements (e.g. from satellites) of relevant
quantities are restricted to the surface. A few previous
results for Lagrangian eddies in 3d were obtained in Refs.
\cite{Branicki2010b,Branicki2011}, by applying the methodology
of lobe dynamics and the turnstile mechanism. Also, Refs.
\cite{Ozgokmen2011,Bettencourt2012} used 3d FSLE fields to
identify LCS in oceanic flows. In particular, a mesoscale eddy
in the Southern Atlantic was studied in \cite{Bettencourt2012},
and it was shown that oceanic LCSs presented a vertical
curtain-like shape, i.e. they look mostly like vertical sheets,
and that material transport into and out of the mesoscale eddy
occurred through filamentary deformation of such
structures.

In this paper, we use 3d fields of FSLE to identify LCSs in a
turbulent channel flow and in an oceanic flow. Observations of
the similarities and differences between the two systems, both
in their computation and their physical meaning, helps to
appreciate the power and scope of this Lagrangian technique in
the analysis of fluid flows. In Section \ref{sec:methods} we
describe the methodology used to identify LCSs in 3d turbulent
flows. Sections \ref{sec:channel_flow} and \ref{sec:ocean_flow}
are devoted to the turbulent channel flow and the oceanic flow,
respectively, and Section \ref{sec:conclus} presents
our conclusions and directions for future work.

\section{Methods}
\label{sec:methods}

\subsection{Finite-Size Lyapunov Exponents.}
\label{subsec:fsle}

In order to study non-asymptotic dispersion processes such as
stretching at finite scales and bounded domains, the finite
size Lyapunov Exponent was introduced \cite{Artale1997,
Aurell1997,Boffetta2001}. It is defined as:
 \begin{equation}\label{FSLE_1}
    \lambda=\frac{1}{\tau}\log\frac{d_{f}}{d_{0}},
 \end{equation}
where $\tau$ is the time it takes for the separation between
two particles, initially $d_{0}$, to reach a value $d_{f}$. 
In
addition to the dependence on the values of $d_0$ and $d_f$,
the FSLE depends also on the initial position of the particles
and on the time of deployment. Locations (i.e. initial
positions) leading to high values of this Lyapunov field
identify regions of strong separation between particles, i.e.,
regions that will exhibit strong stretching during evolution,
that can be identified with the LCS
\cite{Boffetta2001,Joseph2002,dOvidio2004}.

In principle, to compute FSLE in 3d, the method of
\cite{dOvidio2004} can be extended to include the third
dimension, by computing the time it takes for particles
initially separated by $d_0=[(\Delta x_0)^2+ (\Delta
y_0)^2+(\Delta z_0)^2]^{1/2}$ to reach a final distance of
$d_f=[(\Delta x_f)^2+ (\Delta y_f)^2+(\Delta z_f)^2]^{1/2}$. We
will proceed this way for the turbulent channel, but, as
indicated in \cite{Bettencourt2012}, vertical displacements are
much smaller than horizontal ones in ocean flows. Therefore,
the displacement in the $z$ direction does not contribute
significatively to the calculation of $d_f$ in the ocean, which
prompt us to implement a quasi-3d computation of FSLEs: we use
the full 3d velocity field for particle advection but particles
are initialized in 2d horizontal ocean layers and the
contribution $\Delta z_f$ is not considered when computing
$d_f$ (see more details in \cite{Bettencourt2012}). In any
case, since we allow the full 3d trajectories of particles, we
take into account the vertical dynamics of the oceanic flows.

\begin{center}
\begin{figure}[ht]
\centerline{\includegraphics[width=0.5\textwidth]{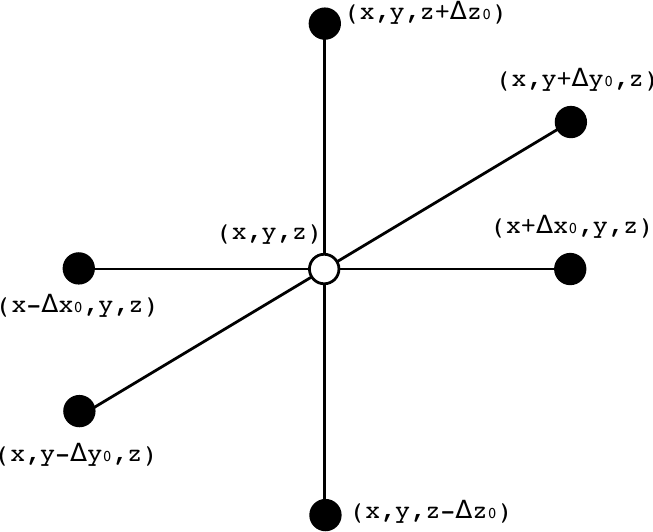}}
\caption{
Computational setup for the calculation of the FSLE field in 3d. The FSLE at the location
of the central particle (\textopenbullet) is a measure of the time it takes for any of the neighbor
particles (\textbullet) to diverge from the central particle by a distance greater than $\delta_f$.
}
\label{Fig1}
\end{figure}
\end{center}

Concerning the turbulent channel, where we can implement a
fully 3d computation of the FSLE, we proceed as follows. A grid
of initial locations $\textbf{x}_{0}=(x_i,y_j,z_k)$ is set up
at time $t$, fixing the spatial resolution of the FSLE field
(figure \ref{Fig1}). Particles are released from each grid point
and their three-dimensional trajectories are calculated. The
distances of each neighbor particle with respect to the central
one (initially $d_0$) is monitored until one of the separations
reaches a value $d_{f}$.

In both systems considered, we obtain two different types of
FSLE maps by integrating the three-dimensional particle
trajectories backward and forward in time: the attracting LCSs
(for the backward), and the repelling LCSs (forward)
\cite{Joseph2002,dOvidio2004}. We obtain in this way FSLE
fields with a spatial resolution given by $d_0$.  When a
particle leaves the velocity field domain or reaches a no-slip
boundary, the FSLE value at its initial position and initial
time is set to zero. If the interparticle separation remains
smaller than $d_{f}$ past a maximum integration time
$\Delta t$, then the FSLE for that location is also set to
zero.

\subsection{Lagrangian Coherent Structures.}
\label{LCS}

The identification of LCS calculated from Lyapunov fields 
in 2d flows is straightforward since they 
practically coincide with (finite-time) stable and
unstable manifolds of relevant hyperbolic structures in the
flow \cite{Haller2000,Haller2000b,Joseph2002} (but see
\cite{BeronVera2010,Haller2011}). The structure of these
 manifolds in 3d is generally much more complex than in 2d
 \cite{Haller2001,Pouransari2010}, and they can be locally
either lines or surfaces.

Differently than 2d, where LCS can be visually identified as
the maxima of the FSLE field, in 3d they are hidden within 
the volume data and one needs to explicitly compute
and extract them, using the definition of LCSs as the ridges of
the FSLE field. 
A ridge $L$ is a co-dimension 1 orientable,
differentiable manifold (which means that for a
3d domain $D$, ridges are surfaces) satisfying
the following conditions \cite{Lekien2007}:
\begin{enumerate}
 \item The field $\lambda$ attains a local extremum at $L$.
 \item The direction perpendicular to the ridge is the
     direction of fastest descent of $\lambda$ at $L$.
\end{enumerate}
The method used to extract the ridges from the scalar field
$\lambda(\mathbf{x}_0,t)$ is from \cite{Schultz2010}. It uses
an earlier \cite{Eberly1994} definition of ridge in the
context of image analysis, as a generalized local maxima of
scalar fields. For a scalar field $f:\mathbb{R}^n \rightarrow
\mathbb{R}$ with gradient $\mathbf{g}=\nabla f$ and Hessian
$\mathbf{H}$, a \textit{d}-dimensional height ridge is given by
the conditions

 \begin{equation}\label{RIDG1}
  \forall {d<i\leq n,} \quad \mathbf{g}^\mathrm{T}\mathbf{e}_i=0 \ \textrm{and}\  \alpha_i<0,  
\end{equation}

where $\alpha_i, i \in \lbrace 1, 2, \ldots, n \rbrace$, are
the eigenvalues of $\mathbf{H}$, ordered such that $\alpha_1
\geq \ldots \geq \alpha_n$, and $\mathbf{e}_i$ is the
eigenvector of $\mathbf{H}$ associated with $\alpha_i$. For
$n=3$, Eq. (\ref{RIDG1}) becomes

\begin{equation}\label{RIDG2}
    \mathbf{g}^\mathrm{T}\mathbf{e}_3=0 \ \textrm{and}\   \alpha_3<0.
\end{equation}

In other words, in $\mathbb{R}^3$ the
$\mathbf{e}_1, \mathbf{e}_2$ eigenvectors point locally along the ridge
and the $\mathbf{e}_3$ eigenvector is orthogonal to it, so the ridge
maximizes the scalar field in the normal direction to it and in this direction
the field is more convex than in any other direction, since the eigenvector
associated with the most negative eigenvalue is oriented along the direction
of maximum negative curvature of the scalar field. 

The extraction process progresses by calculating the points where the
ridge conditions are verified and the ridge strength $|\alpha_3|$ is higher than
a predefined threshold $s$ so that ridge points whose
value of $\alpha_3$ is lower (in absolute value) than $s$ are
discarded from the extraction process.
Since the ridges are
constructed by triangulations of the set of extracted ridge
points, the strength threshold greatly determines the size and shape
of the extracted ridge, by filtering out regions of the ridge
that have low strength.  The reader is referred to
\cite{Schultz2010} for details about the ridge extraction
method. The height ridge definition has been used to extract
LCS from FTLE fields in several works (see, among others,
\cite{Sadlo2007}).

Since the $\lambda$ value of a point on the ridge and the ridge
strength $\alpha_3$ are only related through the expressions
(\ref{RIDG1}) and (\ref{RIDG2}), the relationship between the
two quantities is not direct, which makes difficult to choose
the appropriate strength threshold $s$.
A too small value of
$s$ will result in the extraction of very small LCSs
that appear to have little
influence on the dynamics, while a large value will result in
only a partial rendering of the larger and more significant LCS,
limiting the possibility
of observing their real impact on the flow.

The ridges extracted from the backward FSLE map approximate the
attracting LCSs, and the ridges extracted from the forward FSLE
map approximate the repelling LCSs. The attracting ones are the
more interesting from a physical point of view
\cite{dOvidio2004,dOvidio2009}, since particles (or any passive
scalar driven by the flow) typically approach them and spread
along them, so that they are good candidates to be identified
with the typical filamentary structures observed in tracer
advection.

\section{Turbulent channel flow}
\label{sec:channel_flow}

Turbulent channel flow is a turbulent flow between two stationary,
parallel walls separated by a distance $2 \delta$. It has been
studied extensively due to its geometrical simplicity and its
wall-bounded nature, which makes it a suitable platform to study
phenomena appearing in more complex turbulent wall-bounded
flows of great technological interest.

The coordinates of the flow are: $x$ for the streamwise
direction, $y$ for the cross-stream coordinate that separates
the two plates, and $z$ for the spanwise direction. The flow is
maintained by a downstream pressure gradient $\frac{dP_0}{dx}$
acting against the wall shear stress. The laminar flow solution
$U_0$ is a cross-stream parabolic profile given by
\begin{equation}
 U_0(y)=\frac{y^{2}-\delta^{2}}{2\mu}\frac{dP_0}{dx} \ ,
\end{equation}
where $\mu$ is the dynamic viscosity. Following the Reynolds
averaging method \cite{Tennekes72}, the turbulent flow velocity
$\mathbf{u}$ is decomposed in a mean  $\mathbf{U}= (U(y),0,0)$
and a fluctuating component $\mathbf{u}'=(u',v',w')$. The mean
turbulent velocity profile $U(y)$ differs from the laminar one,
$U_0(y)$, by a lower centerline velocity $U(0)$ and increased
near-wall velocity giving it a flatter shape. Due to the
increase in mean velocity near the wall, the shear stress near
the wall is higher for the turbulent case. The total shear
stress $\tau$ appearing in the averaged Reynolds equations gets
contributions from both the viscous stress and the Reynolds
stress $-\overline{u'v'}$ associated to the velocity
fluctuations:
\begin{equation}
 \frac{\tau}{\rho}=\nu\frac{dU}{dy}-\overline{u'v'}
\end{equation}
$\nu={\mu}/{\rho}$ is the kinematic viscosity. The symmetries
of the domain and the Reynolds equations imply that $\tau$
depends only on the cross-stream coordinate $y$, and the
dependence is linear, so that it can be written as
\begin{equation}
 \frac{\tau(y)}{\rho}=u_{\tau}^2 \left(1-\frac{y}{\delta}\right)
\end{equation}
The shear velocity $u_{\tau}$ gives the velocity scale of the
turbulent velocity fluctuations. The formula \cite{Tennekes72}:
\begin{equation}
\rho u_{\tau}^2 =\mu \left.\frac{dU(y)}{dy}\right|_{y=0}
\label{utau}
\end{equation}
allows to compute $u_\tau$ from measurements of the mean
velocity profile from the simulations. A length scale can be
formed by combining $u_\tau$ with $\nu$: the wall scale
$\delta^{+}={\nu}/{u_{\tau}}$. The wall distance can now be
expressed as $y^{+}=y/\delta^{+}$, and the same normalization
could be done for the rest of coordinates. The viscous Reynolds
number $Re_{\tau}={\delta}/{\delta^{+}}$ is simply the ratio
between the two relevant length scales.

The existence of coherent structures in turbulent wall-bounded
flows has been known for several decades from investigations on
intermittency in the interface between turbulent and potential
flow regions, on the large eddy motions in the outer regions of
the boundary layer, and on coherent features in the near-wall
region (\cite{Robinson1991} and references therein). Since
then, through experimental and numerical investigations, a
picture of the organization of these coherent structures in the
turbulent boundary layer has emerged, which has become rather
complete from the Eulerian point of view
\cite{Robinson1991,Holmes1998}. Our approach is a contribution
to the Lagrangian exploration of these coherent structures, as
in \cite{Green2007} and \cite{Pan2009}.

The longitudinal velocity field in the inner region of the
channel (the viscous sublayer adjacent to the wall and the
intermediate buffer region) is organized into alternating
streamwise streaks of high and low speed fluid. Turbulence
production occurs mainly in the buffer region in association
with intermittent and violent outward ejections of low-speed
fluid and inrushes of high-speed fluid towards the wall. The
outer region is characterized by the existence of
three-dimensional $\delta$-scale bulges that form on the
turbulent/potential flows interface. Irrotational valleys
appear at the edges of the bulges, entraining high-speed fluid
into the turbulent inner region. A central element in the
structure of the turbulent boundary layer is the hairpin
vortex, mainly because it is a structure with the capability of
transporting mass and momentum across the mean velocity
gradient and because it provides a paradigm with which to
explain several observations of wall turbulence
\cite{Robinson1991,Adrian2007}.

\subsection{Data}
\label{cf_data}

The data used to extract the LCS come from a direct numerical
simulation (DNS) of turbulent channel flow at a viscous
Reynolds number $Re_{\tau}=180$. The setup of the simulation
follows that of \cite{mkm99} and is
summarized in table \ref{Tab1}. The simulations were conducted
using the CFD solver \texttt{Channelflow.org} \cite{channelflow}. 
The \texttt{Channelflow.org} code
solves the incompressible Navier-Stokes equations in a
rectangular box with dimensions $L_x \times 2\delta \times
L_z$, with periodic boundary conditions in the $x$ (so that
fluid leaving the computational domain in the direction of the
mean flow at $x=L_x$ reenters it at $x=0$) and in the spanwise
$z$ direction. No-slip conditions are imposed on $y=\pm
\delta$. The unsteady velocity field $\bf{u}$ is represented as
a combination of Fourier modes in the $x$ and $z$ directions
and of Chebyshev polynomials in the wall-normal direction. The
pressure gradient necessary to balance the friction at the
walls was chosen as to maintain a constant bulk velocity of
$\frac{2}{3} U_0$. Time stepping is a 3rd-order Semi-implicit
Backward Differentiation. Note that in our computations
$\delta^+=1/Re_{\tau}=0.0058$ so that in wall units $0 < y^+ <
344$.

The flow was integrated from an initial base-flow with
parabolic profile and a small disturbance that evolved into a
fully developed turbulent flow. The total integration time was
$\Delta t=600$ time units that in dimensionless form $t^{+}=t
\left(u_{\tau}^{2}/\nu\right)$ gives $\Delta t^{+}=83.54$.
After an initial transient of about $200$ time units the
simulations reached a statistically stationary state from which
statistics was accumulated.

\begin{figure}
\centerline{\includegraphics[width=0.7\textwidth]{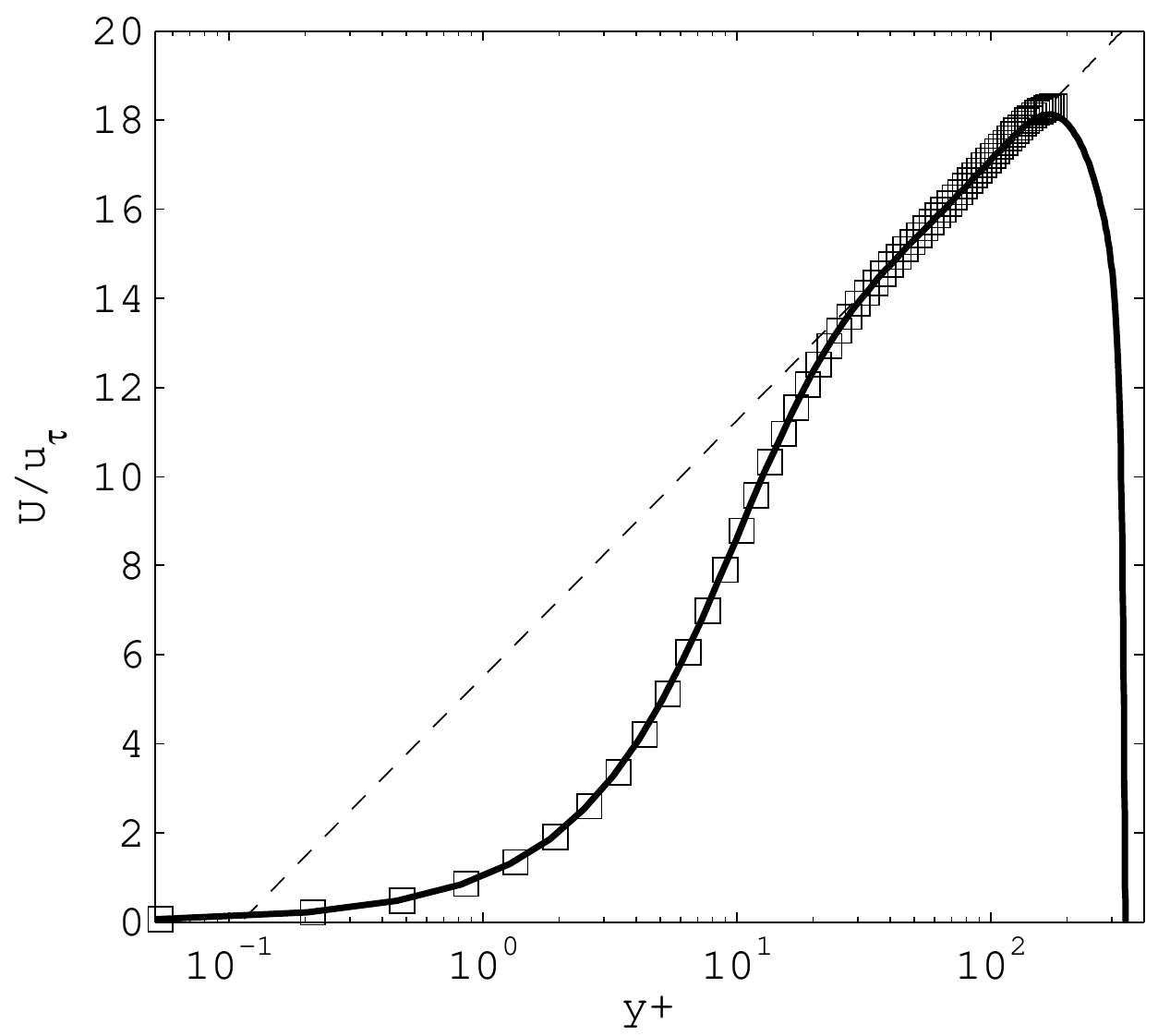}}
\caption{
Mean velocity profile $U(y)/u_{\tau}$ . Solid line: our simulations;
squares: \cite{mkm99}; dashed line: logarithmic profile $U(y)/u_{\tau}=2.5 \log(y^+)+5.5$.
}
\label{Fig2}
\end{figure}

\begin{figure}
\centerline{\includegraphics[width=0.7\textwidth]{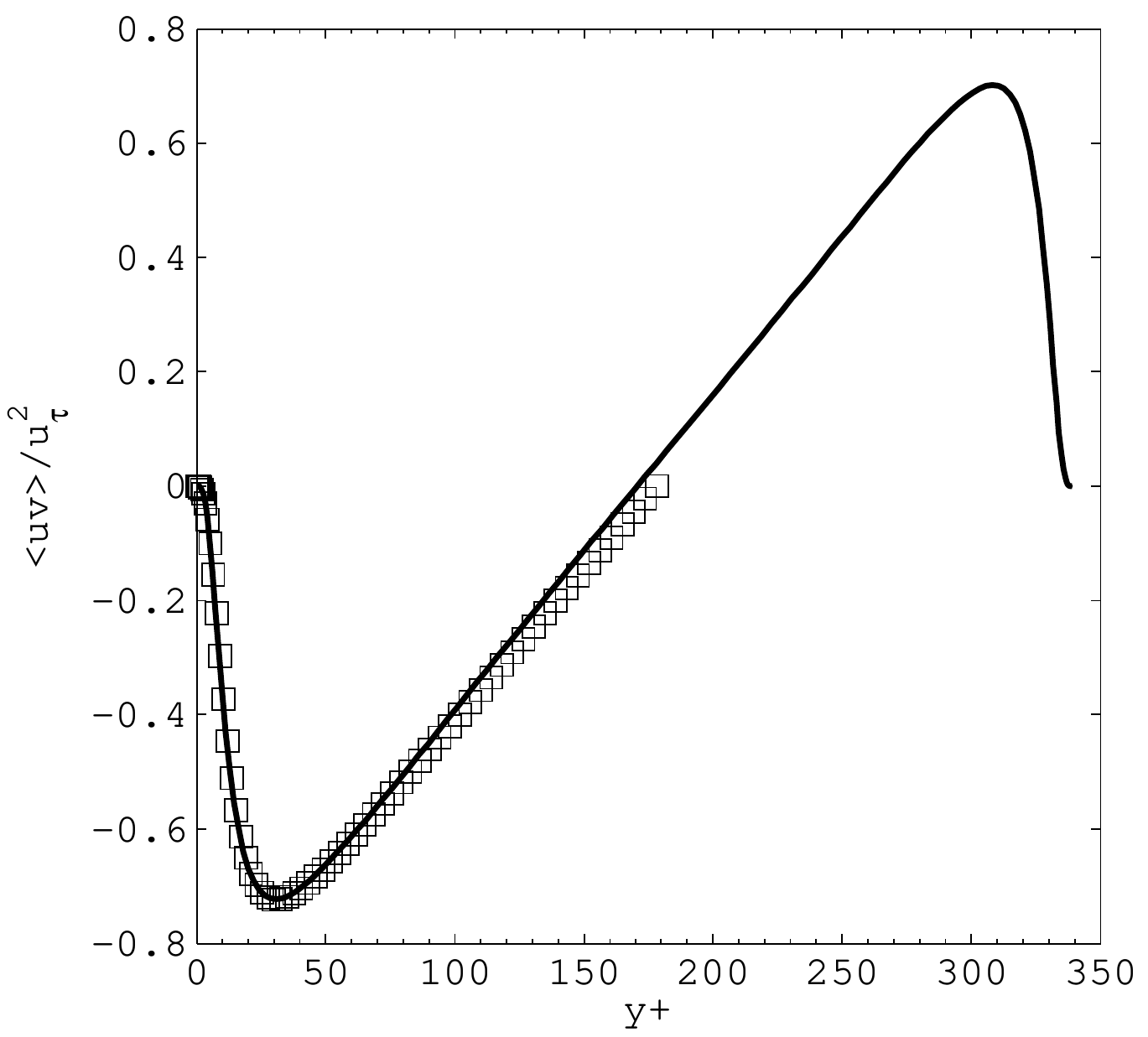}}
\caption{
Reynolds stress $\overline{u'v'}$ profile at $Re_{\tau}=180$. Solid line: our
simulations; squares: \cite{mkm99} (given up to the channel centerline).
}
\label{Fig3}
\end{figure}

The mean quantities and first order statistics of our
simulations where compared to those of \cite{mkm99} and the
agreement is quite good. The profile of the mean velocity in
wall units is shown in figure \ref{Fig2}.
The profile for the Reynolds stress $-\overline{u'v'}$ shows
that the maximum (in absolute value) is located at
approximately $y^{+}=30$, in the outer limit of the buffer
layer (see figure \ref{Fig3}).

\begin{table}
\caption{\label{Tab1}Simulation parameters. Quantities with
$^+$ refer to wall units. $L_x$, $2\delta$ and $L_z$ are the
domain sizes in the $x$, $y$ and $z$ directions. $\Delta x^+$,
$\Delta y^+$ and $\Delta z^+$ are the respective spatial
resolutions (given at the first point above the wall for the
$y$ case), and $n_x$, $n_y$ and $n_z$ the corresponding number
of grid points. $Re=U\delta/\nu$ is the Reynolds number based
on the channel center mean speed, whereas
$Re_\tau=u_\tau\delta/\nu$ is the viscous Reynolds number. The
nominal value is an input to the computer code, and
the actual value comes by using Eq. (\ref{utau}) for the
computed mean profile $U(y)$. }
\begin{indented}
\item[]\begin{tabular}{@{}ll|ll|ll} \br
$Re$ channel center &$4000$     &$Re_{\tau}$ nominal &$180$    &$Re_{\tau}$ actual &$172$\\
$L_x$               &$4\pi$     &$\delta$            &$1$      &$L_z$              &$\frac{4}{3}\pi$ \\
$L_{x}^+$           &$2166.61$  &$\delta^+$	         &$0.0058$ &$L_{z}^+$          &$722.20$\\
$n_x$               &128        &$n_y$               &129      &$n_z$              &128\\
$\Delta x^+$        &$17.06$    &$\Delta y^+$        &$0.005$  &$\Delta z^+$       &$5.6867$\\
\br
\end{tabular}
\end{indented}
\end{table}

\subsection{Results}

The LCS were extracted from the turbulent velocity field data
described in the previous section. A calculation of FSLE field
in the entire turbulent channel was conducted in order to
understand the statistical properties of the FSLE field in this
class of turbulent flows. A subsequent calculation in a
subdomain of the channel was used to extract the LCS in that
subdomain for a sequence of time instants. The setup of both
calculations is shown in table \ref{Tab2}.

\begin{table}
\caption{\label{Tab2}FSLE calculation parameters. $dt$ is the
integration time step and $\Delta t$ the maximum integration
time.}
\begin{indented}
\item[]\begin{tabular}{@{}lllll}
\br
Calculation      &$d_0$ &$d_f /d_0$ &$\Delta t$ &$dt$\\
\hline
\hline
Complete channel &$0.024$    &$20$                  &$20$          &$0.05$\\
LCS subdomain    &$0.003$    &$67$                  &$10$          &$0.05$\\
\br
\end{tabular}
\end{indented}
\end{table}

\subsubsection{The 3d FSLE field.}
\label{3d_fsle_cf}

The 3d backward FSLE field for the entire channel was
calculated at a single time instant in the statistically steady
state. The initial and final distances $d_0$ and $d_f$ were
chosen as a balance between encompassing the widest possible
range of scales of motion (measured by the ratio $d_f /d_0 $),
and adequate resolution and computational cost. The initial
distance is of the order of $4 \delta^{+}$ and the final
distance of the order of $0.5 \delta$ -- a typical scale of coherent structures found
in the turbulent channel flow -- so that the ratio of
scales, $d_f /d_0$, is approximately $Re_{\tau}/8$.

\begin{figure}[ht]
\centerline{\includegraphics[width=.9\textwidth, height=7cm]{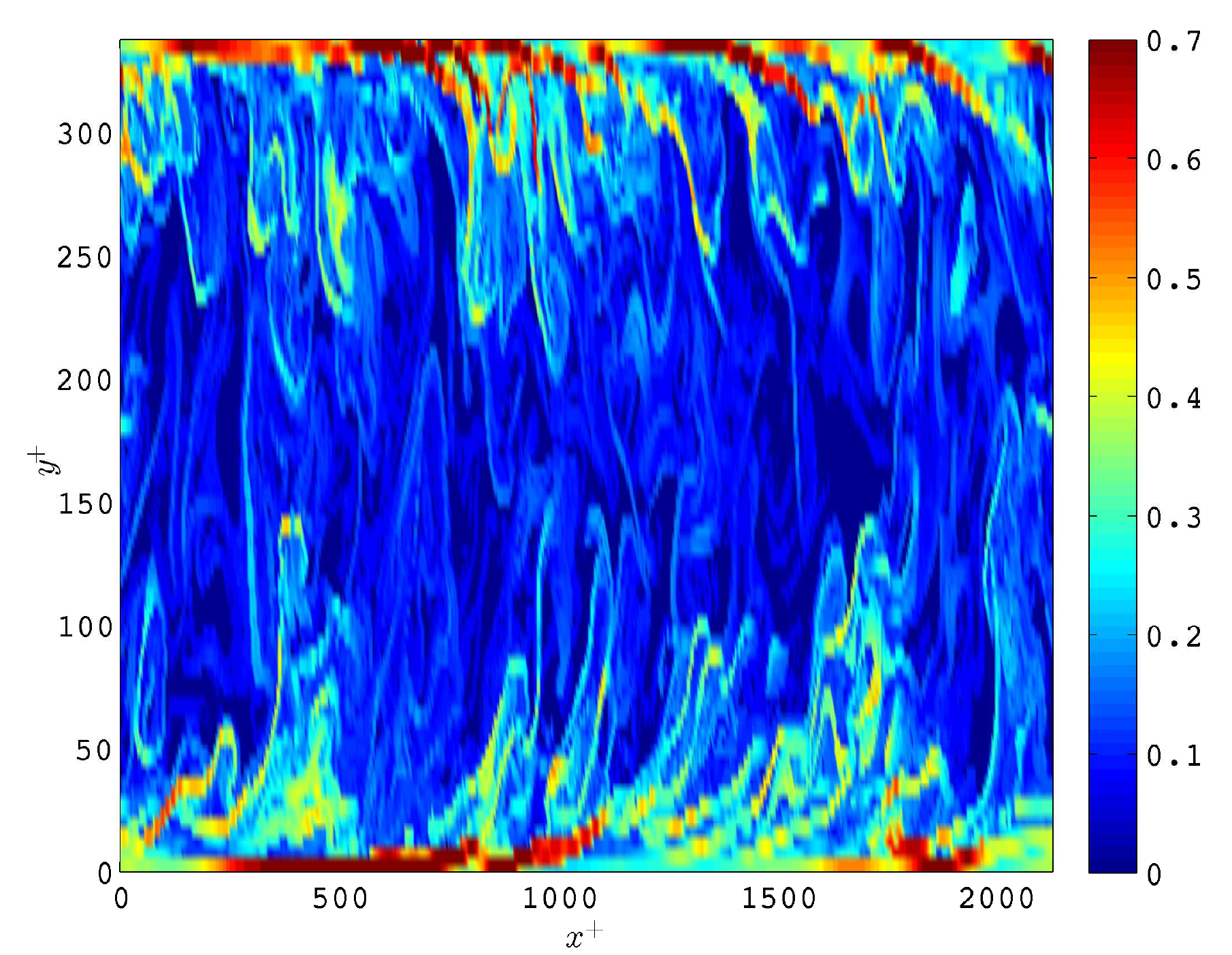}}
\caption{
Instantaneous FSLE field at $t=420$ shown on a streamwise/wall-normal plane in the turbulent channel. Walls are at the
top and bottom of the figure. Mean velocity is in the streamwise direction 
from left to right.
}
\label{Fig4}
\end{figure}

\begin{center}
\begin{figure}
\centerline{\includegraphics[width=0.7\textwidth]{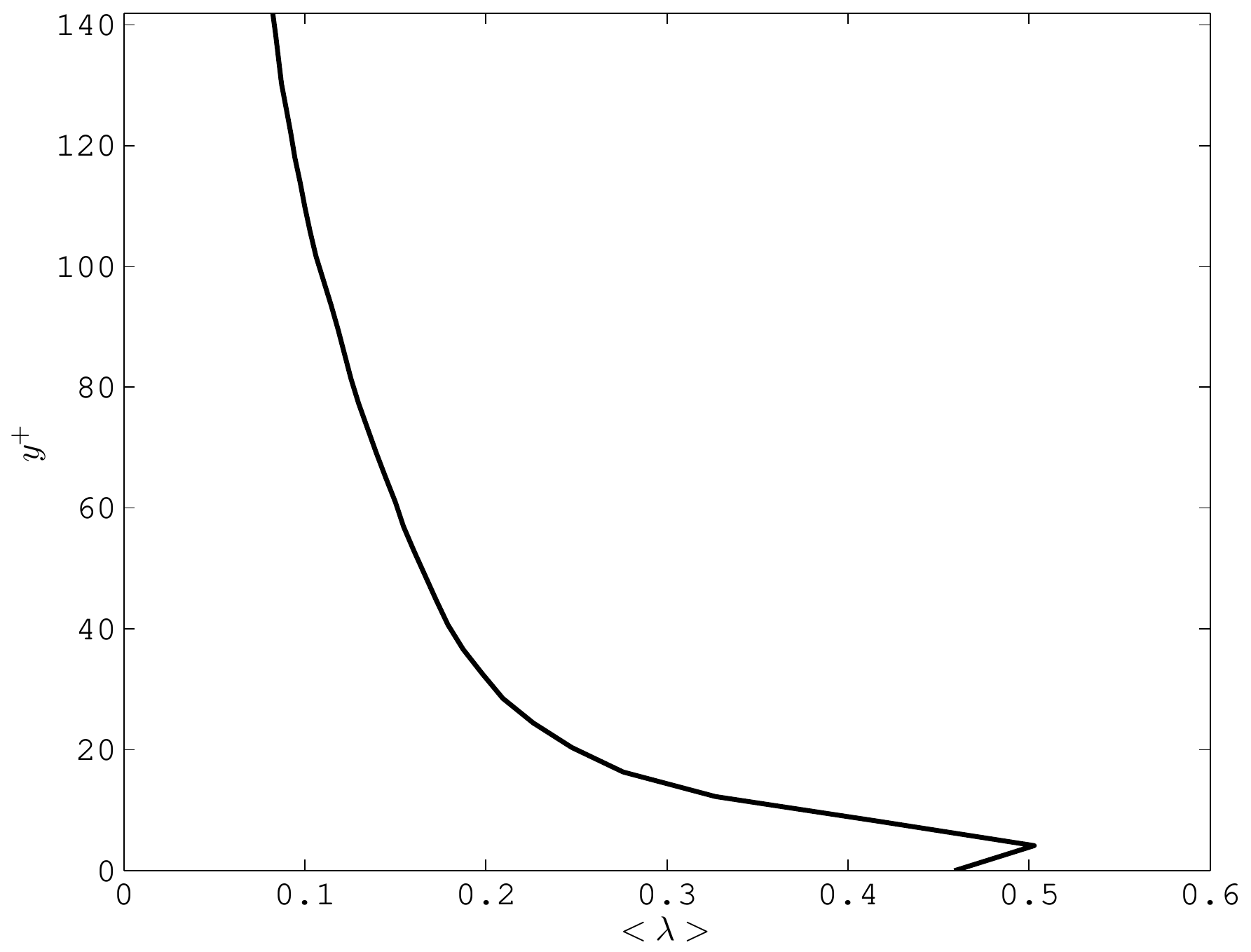}}
\caption{
FSLE profile averaged over $(x,z)$, as a function of the cross-stream
normalized coordinate $y^+$. Only half of the channel is shown since
the profile is quasi-symmetric about the channel centerline.
}
\label{Fig5}
\end{figure}
\end{center}

Figure \ref{Fig4} shows an instantaneous configuration of the
FSLE values in a streamwise/wall-normal plane. 
The maxima of the FSLE appear to be located close to the walls
with ocasional sloping structures extending to the midchannel 
region. The channel center is devoid of
high FSLE values but coherent patches of low FSLE values can
still be observed. These structures are not distributed uniformly
along the length of the channel but appear to be organized in
packets. This organization bears
resemblance to the widely accepted picture of organized
structures in wall turbulence where the outer region is
dominated by packets of sloping hairpin vortices and the inner
region by near wall vortices (the hairpin vortices legs) and
shear layers \cite{Adrian2007,Robinson1991}. 

A cross-stream FSLE profile is obtained by averaging the 3d
field over the periodic directions $x$ and $z$. It is shown in
figure \ref{Fig5}. The profile is symmetric about the channel
centerline and shows a maximum at approximately $y^{+}=4$, 
inside the viscous sublayer (this location corresponds to the
first grid point off the wall). 

\begin{center}
\begin{figure}
\centerline{\includegraphics[width=0.7\textwidth]{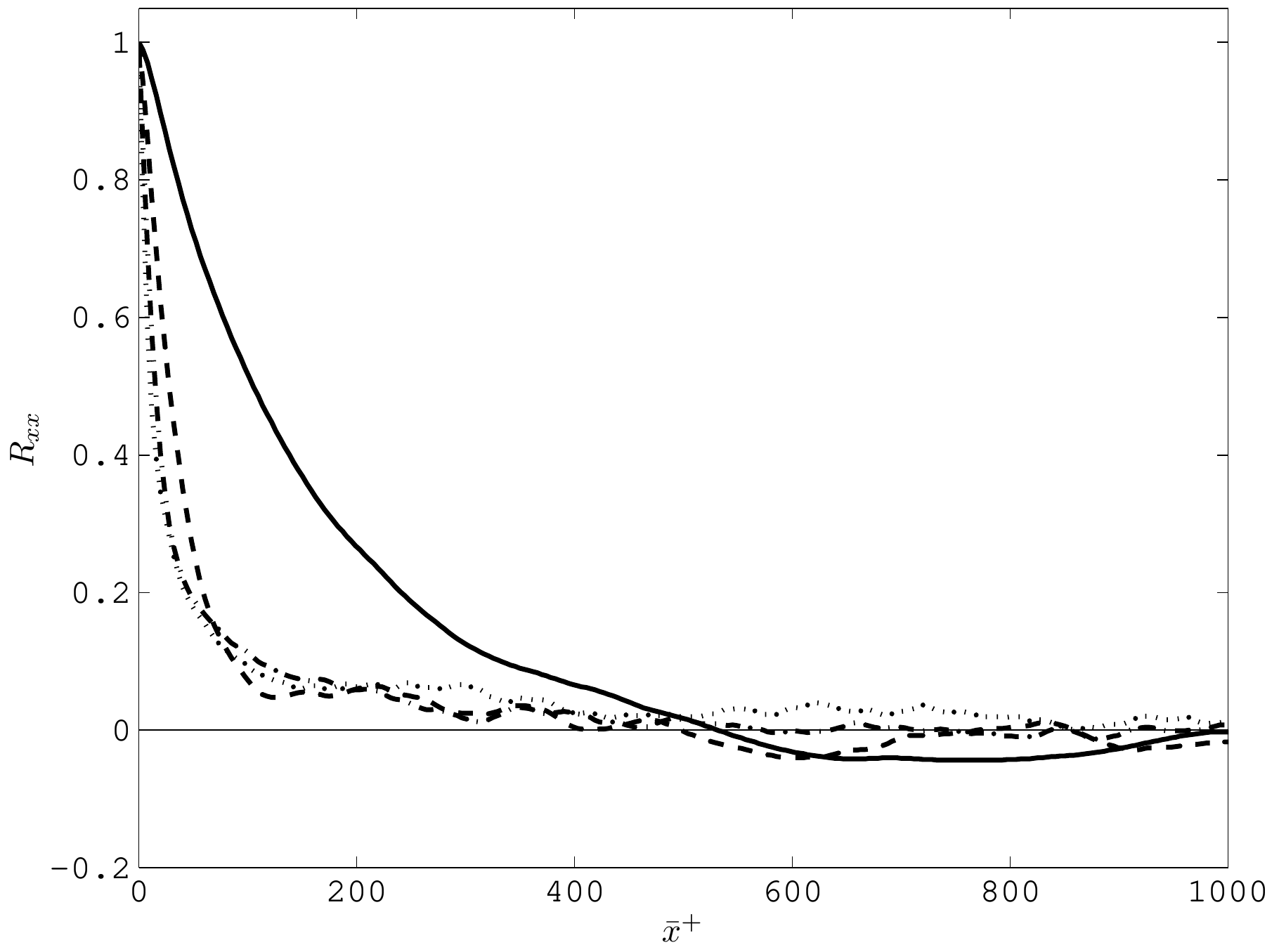}}
\caption{
Streamwise correlation function $R_{xx}(y^+; \bar{x}^+)$ as a function of the
streamwise separation $\bar{x}^+$, at four distances from the lower wall: Continuous line: $y^+=4$; 
dashed line $y^+=12.2$; dash-dot line $y^+=28.4$; 
dotted line: $y^+=122.1$.
}
\label{Fig6}
\end{figure}
\end{center}

\begin{center}
\begin{figure}
\centerline{\includegraphics[width=0.7\textwidth]{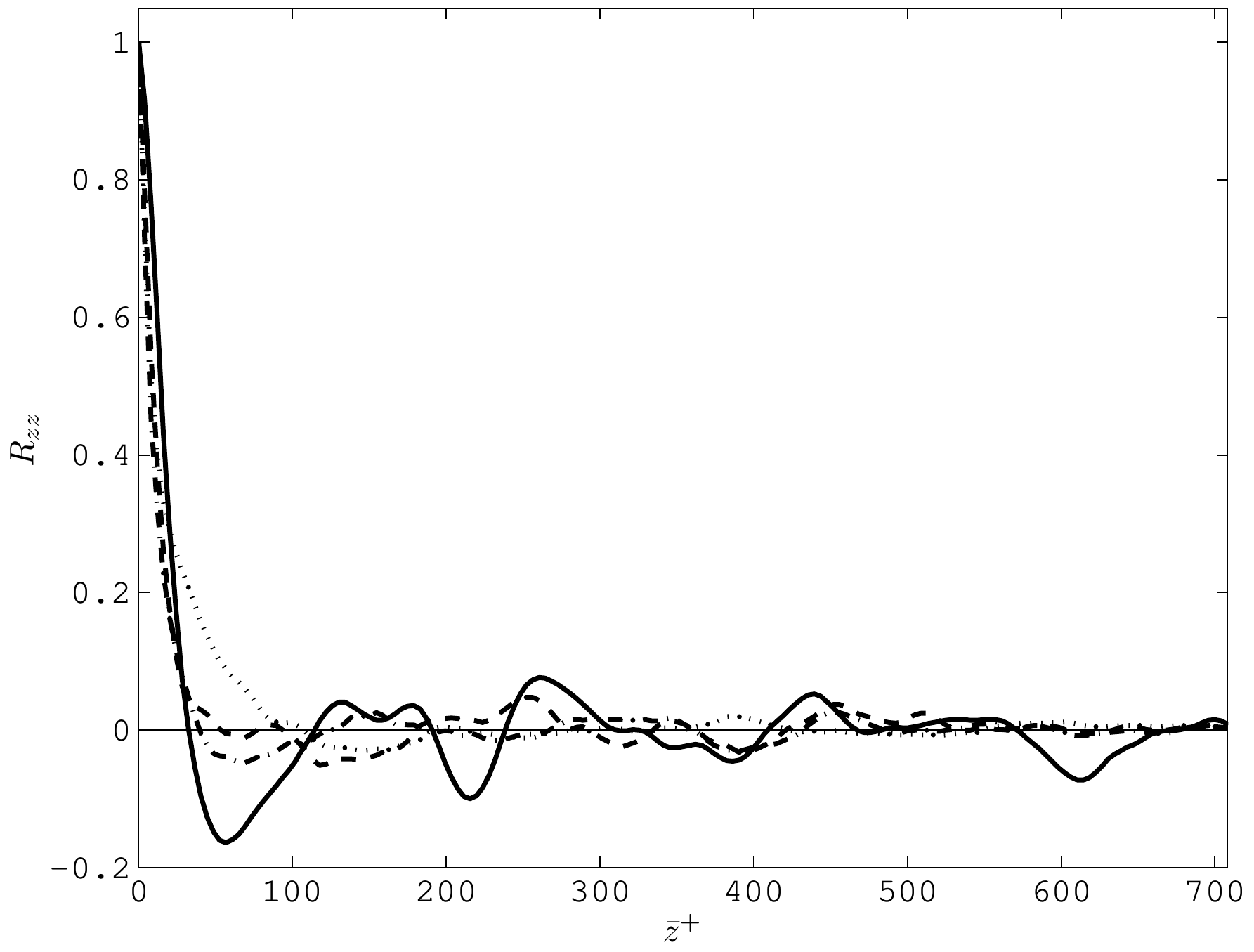}}
\caption{
Spanwise correlation function $R_{zz}(y^+; \bar{z}^+)$ as a function of the
spanwise separation $\bar{z}^+$, at four distances from the lower wall: Continuous line: $y^+=4$; 
dashed line $y^+=12.2$; dash-dot line $y^+=28.4$; 
dotted line: $y^+=122.1$.}
\label{Fig7}
\end{figure}
\end{center}

Because of the periodic boundary conditions
in the $x$ and $z$ directions the average profiles along these
directions are rather unstructured, and we resort to two-point
correlation functions to quantify the statistical structure
properties. For each plane parallel to the walls, i.e. for each
value of $y^+$, we compute the fluctuations of the FSLE values
around the average in that plane: $\Lambda(x^+,y^+,z^+) \equiv
\lambda(x^+,y^+,z^+)-\left<\lambda(x^+,y^+,z^+)\right>_{x^+,z^+}$.
From this quantity we define the streamwise
$R_{xx}(y^+;\bar{x}^+)$ correlation function as:
\begin{equation}
R_{xx}(y^+; \bar{x}^+)=\frac{
\left<\Lambda(x^+,y^+,z^+)\Lambda(x^++\bar{x}^+,y^+,z^+)\right>_{x^+,z^+}
}{
\left<\Lambda(x^+,y^+,z^+)^2\right>_{x^+,z^+}
} \ ,
\end{equation}
and the spanwise $R_{zz}(y^+;\bar{z}^+)$ correlation function
\begin{equation}
R_{zz}(y^+; \bar{z}^+)=\frac{
\left<\Lambda(x^+,y^+,z^+)\Lambda(x^+,y^+,z^++\bar{z}^+)\right>_{x^+,z^+}
}{
\left<\Lambda(x^+,y^+,z^+)^2\right>_{x^+,z^+}
} \ .
\end{equation}
In the above equations the averages are over the periodic
directions $x^+$ and $z^+$. The correlations are shown in Figs.
\ref{Fig6} and \ref{Fig7} at different distances from the
walls: one smaller, one larger, and one approximately
coincident with the location of the maximum Reynolds stress.
These functions reveal sizes and organization of the different
structures in the Lagrangian FSLE field, to be contrasted with
Eulerian correlation functions in the same system
\cite{kmm87}.

Close to the wall ($y^+=4$ and $y^+=12.2$), viscous effects 
dominate. The correlations show that the FSLE field is organized 
in streamwise structures of length scale approximately $l^+ \sim 500$
wall units. 
In the transverse direction $z^+$
the oscillations seen in $R_{zz}$ for $y^+=4$ indicate an
approximately periodic arrangement of the streaks
\cite{Green2007}, with a spacing $\sim 50 - 100$ wall
units. This pattern of organization is similar to what is seen
in Eulerian descriptions \cite{kmm87,Robinson1991}. 

At planes further away from the wall ($y^+=28.4$ and $y^+=122.1$ in Figs.
\ref{Fig6} and \ref{Fig7}), correlation functions in both
directions become shorter ranged, and periodic features are
progressively lost. This corresponds to a rather disordered
distribution of structures, each with a typical size related to
the width of the correlation functions, i.e. of the order of 50
wall units, as also seen in figure \ref{Fig4}.

\begin{center}
\begin{figure}
\centerline{\includegraphics[width=0.9\textwidth, height=7cm]{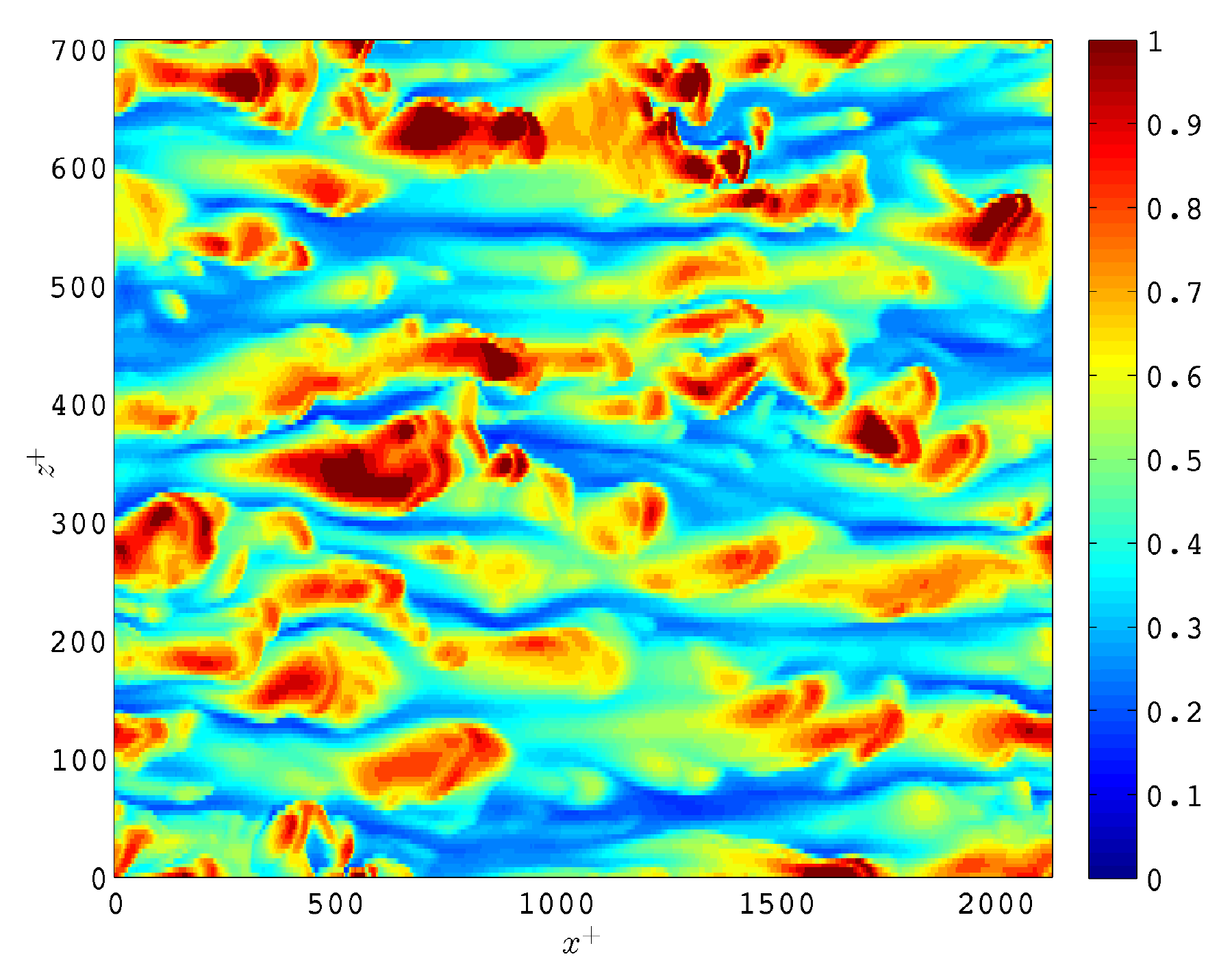}}
\caption{
Instantaneous FSLE field in plane parallel to the wall at $y^+=4$. 
The time is the same as in
figure \ref{Fig4} }
\label{Fig8}
\end{figure}
\end{center}

An instantaneous near-wall FSLE field is shown in figure \ref{Fig8},
where the high FSLE values appear in slender and elongated structures
with length and width corresponding to the streamwise and spanwise
correlation lengths discussed above. It is unclear whether the correlation
lengths result from a single streamwise structure or from the overlaping
of shorter structures (a feature of the near wall coherent structure
arrangement \cite{Jeong1997}).

\begin{center}
\begin{figure}
\centerline{\includegraphics[scale=0.4]{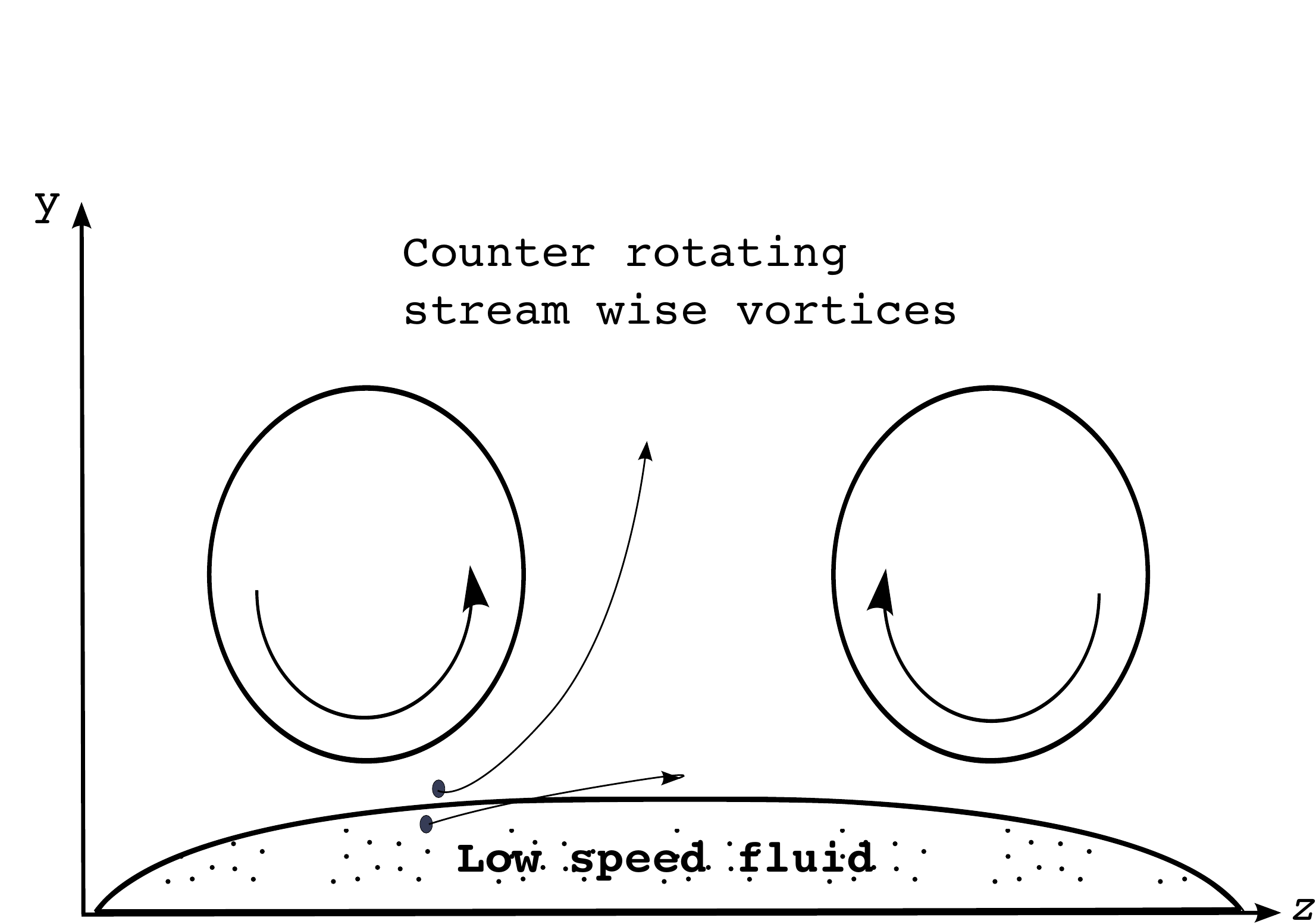}}
\caption{
Mechanism for the rapid separation of fluid from the near wall viscous 
sublayer. The mean flow is into the page.}
\label{Fig9}
\end{figure}
\end{center}

These are the highest FSLE values that are to be found in the channel as 
the plot in figure \ref{Fig5} shows. 
The mechanism for the formation of these structures could be
the lifting of low speed fluid close to the wall by the action of counter
rotating vortex pairs located above the viscous sublayer (see figure \ref{Fig9}). 
This mechanism is widely known in the Eulerian view of coherent structures of turbulent 
wall bounded flows (\emph{ejections} or \emph{bursting}, \cite{Adrian2007}).

The near wall fluid is advected away from the wall by the action of these vortices. 
This mechanism could be responsible for very fast particle separation in 
particle pairs where one particle is lifted away and the other remains in the low
speed zone close to the wall. We note that the particle separation would increase not
only by the wall normal distance between particles but also because the ejected particle
would move to a region with higher streamwise velocity. Shear layers
near the wall is another possible way to produce large particle dispersion.
These mechanisms would explain the fact that the maximum average FSLE 
is located so close to the wall and not on the buffer region where turbulence
production is larger. To conclude, we note that these high FSLE regions near the wall
seem to extend to the midchannel region in an inclined fashion. 
It is not clear whether this pattern signals the existence of a hairpin
vortex with streamwise legs and inclined head or if there are two separate
structures: the streamwise vortices \emph{and} the hairpin arch or head 
\cite{Robinson1991}. Also, we note that the interpretation of the high FSLE
regions near the wall do not require the existence of a counter rotation pair
of vortices, as only one vortex would suffice.

\begin{center}
\begin{figure}
\centerline{\includegraphics[width=0.9\textwidth]{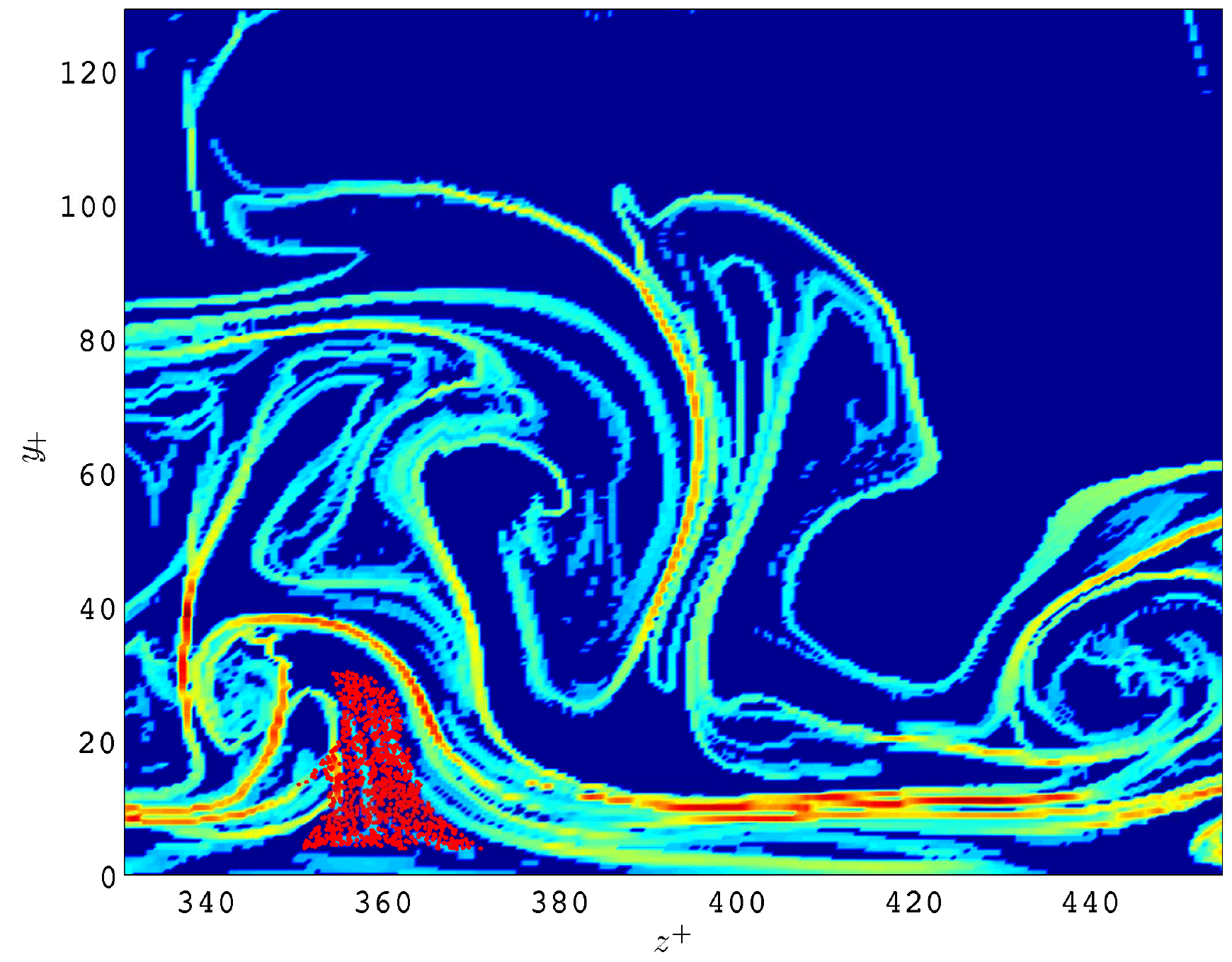}}
\caption{
FSLE map in a $(z,y)$ plane located at $x=6.0$ ($x^+=1034$). The time of the map is $t=413.8$.
together with a set of particles initially located in rectangular region 
$z^+ \in [345, 380]$ and $y^+ \in [3.4, 13.8]$. The particles were released at $t=409$. 
Particle trajectories were integrated using only the spanwise and wall normal velocity
components. The mean flow is moving out of the page.
}
\label{Fig10}
\end{figure}
\end{center}

To illustrate these mechanisms, a map of the FSLE field in a spanwise/wall normal
plane for the LCS domain calculation is shown in figure \ref{Fig10}, together with a
set of passive particles initially located in a rectangular region close to the wall
and released some instants before the time of the FSLE map. In order to focus
just on the above mentioned ejection mechanism involving only
the vertical motion of the particles, the trajectory integration was made in a 2d
fashion by setting the longitudinal component of the particles velocity to zero.

\begin{center}
\begin{figure}
\centerline{\includegraphics[width=0.9\textwidth]{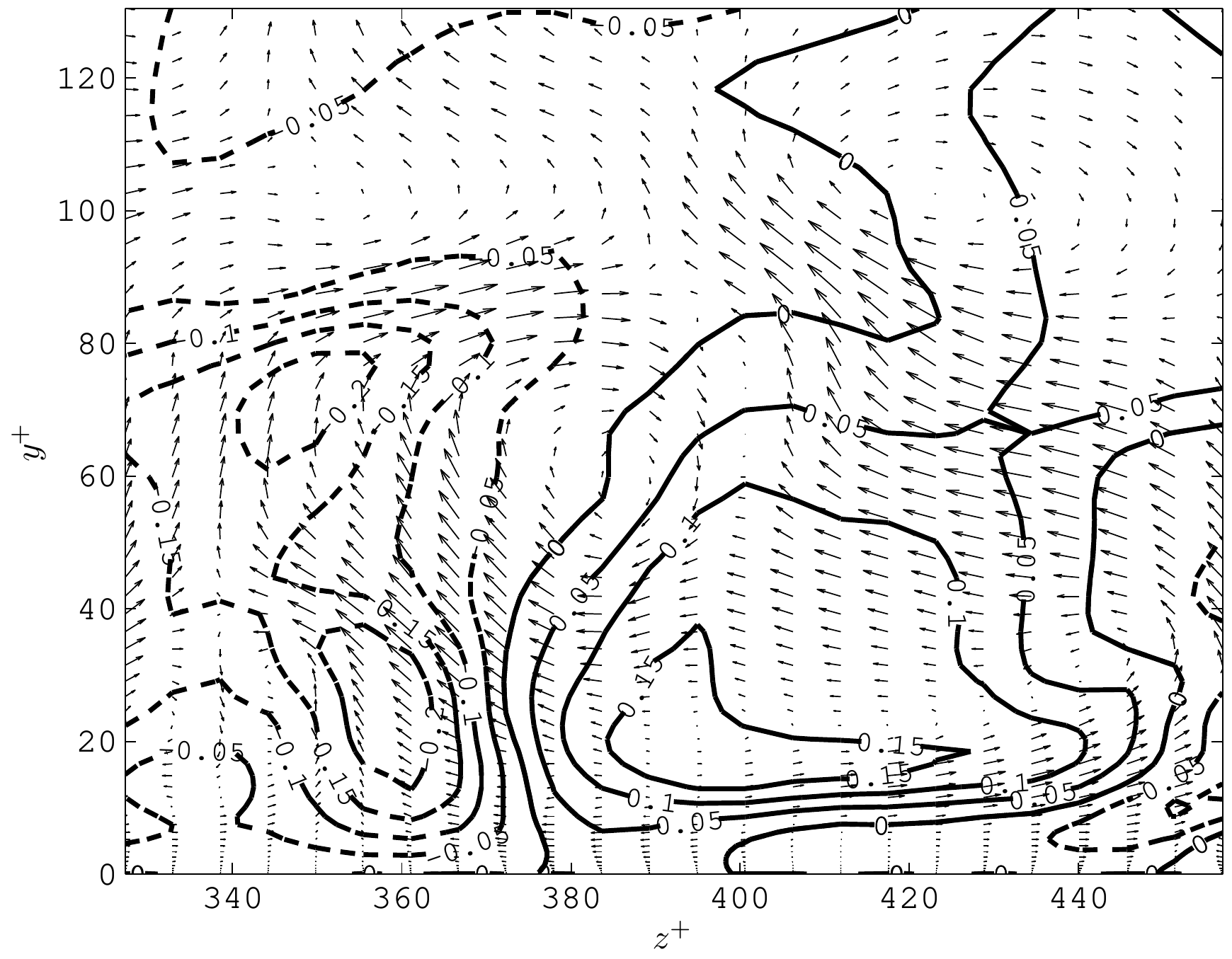}}
\caption{
Instantaneous turbulent velocity components at $x=6.0$ ($x^+=1034$) and $t=413.75$. 
Velocity vectors correspond to the inplane velocity components $(w,v)$,
together with contours of streamwise turbulent velocity $u$. Dashed contours
are negative $u$ (into the paper) and continuous countours are positive $u$
(from the paper). 
}
\label{Fig11}
\end{figure}
\end{center}

The particles seem to have been lifted from wall by a streamwise vortex located
to the left of the particle plume, with center at $(z^+,y^+)\sim (340,30)$. We note that
the structures are moving with the mean flow and that the continuous motion of
the particles away from the wall is due to the passage of a streamwise structure
that imparts this sustained motion to the particles for long enough time. 
To compare the Eulerian and Lagrangian coherent structures, figure \ref{Fig11} 
shows the turbulent velocity components in the same plane
at the nearest time available in the turbulent dataset. The signature 
of the streamwise vortex discussed above can be seen in the Eulerian map 
at the same location. It is embedded in a patch of negative streamwise velocity
fluctuation $u$. To the right, close to $z^+=380$, a vertical shear layer appears
dividing the negative and positive patches of $u$. The Lagrangian signature of
this vertical shear layer is not very strong and appears in figure \ref{Fig10} 
as quasi-vertical line of moderate FSLE extending from $y^+=25$ to $y^+=60$. 
On the lower right of the map, there is a set of high FSLE lines almost parallel
to wall, signalling the existence of high particle dispersion. In the
Eulerian map (figure \ref{Fig11}), it can be seen that there is a shear layer
parallel to the wall at the same location ($400 < z^+ < 440$ and $y^+ \sim 8$).
The fact that this shear layer has a much stronger Lagrangian signature than
the vertical shear layer could be because it has the same orientation and sign
of the mean shear and therefore acts together with the latter to disperse neighboring
particles across the wall normal direction. The high FSLE line seen at
the middle of the map in figure \ref{Fig10}, separating the two convoluted
features can be seen to be related to the existence of two counter-rotating
vortices, one with center located at $\sim (380,60)$ and the other at
$\sim (420, 100)$. The line of high FSLE line is seen to be located at the
boundary between both vortices.
In section \ref{3d_LCS_cf}, we present a 3d view of these structures and 
their evolution in time.

\subsubsection{Propagation velocity.}
\label{pvel_cf}

In turbulent channel flow the velocity perturbations propagate
in the streamwise direction aproximately with the 
velocity of the mean flow\cite{KimHussain1993}.
In the case of Lyapunov exponents, \cite{Pan2009} measured 
the FTLE field in an 2D turbulent boundary layer velocity field obtained 
by time-resolved PIV measurements. The FTLE maxima were found to move 
with the mean flow velocity.

We measured the propagation velocity of the FSLE field perturbation
using a space-time correlation of the form:

\begin{center}
\begin{equation}
R_{uu}(y^+; \bar{x}^+; \bar{t}^+)=\frac{
\left<\Lambda(x^+,y^+,z^+,t^+)\Lambda(x^++\bar{x}^+,y^+,z^+,t^++\bar{t}^+)
\right>_{x^+,z^+}
}{
\left<\Lambda(x^+,y^+,z^+,t^+)^2\right>_{x^+,z^+}
}  \ ,
\label{Ruu}
\end{equation}
\end{center}

where $\bar{x}^+$ and $\bar{t}^+$ are the delays in the streamwise direction
and time. The time delay is fixed and the propagation velocity is defined as

\begin{equation}
 V^+=\frac{\bar{X}^+}{\bar{t}^+} \ ,
\label{vprop}
\end{equation}

where $\bar{X}^+$ is the streamwise lag for which $R_{uu}$ 
is maximum. The choice of the time delay is related to the 
time scale of the FSLE field. A first rule is to choose
a time delay that gives reasonable peaks in the correlation.
If there are several time scales present, several $\bar{t}^+$
will result in correlations exhibiting peaks. 
The calculation of (\ref{vprop}) was made for a full length and height
spanwise section of the channel. A time series of FSLE fields 
with time step of $dt^+=1.8$ and time length $\Delta t^+=431$
was calculated for this section to offset the effects of the
limited spanwise extent of the section. The final time lag used
in (\ref{vprop}) was equal to $dt^+$. All larger delays produced
correlations with no significant peak. A reason for this could
be the fact that by setting the FSLE final distance the length
scales of turbulence retained in the FSLE field is fixed, and 
then there will be only one time delay producing a peak in the
correlation (\ref{Ruu}), specifically that corresponding to 
$V^+$. 

\begin{center}
\begin{figure}
\centerline{\includegraphics[width=0.7\textwidth]{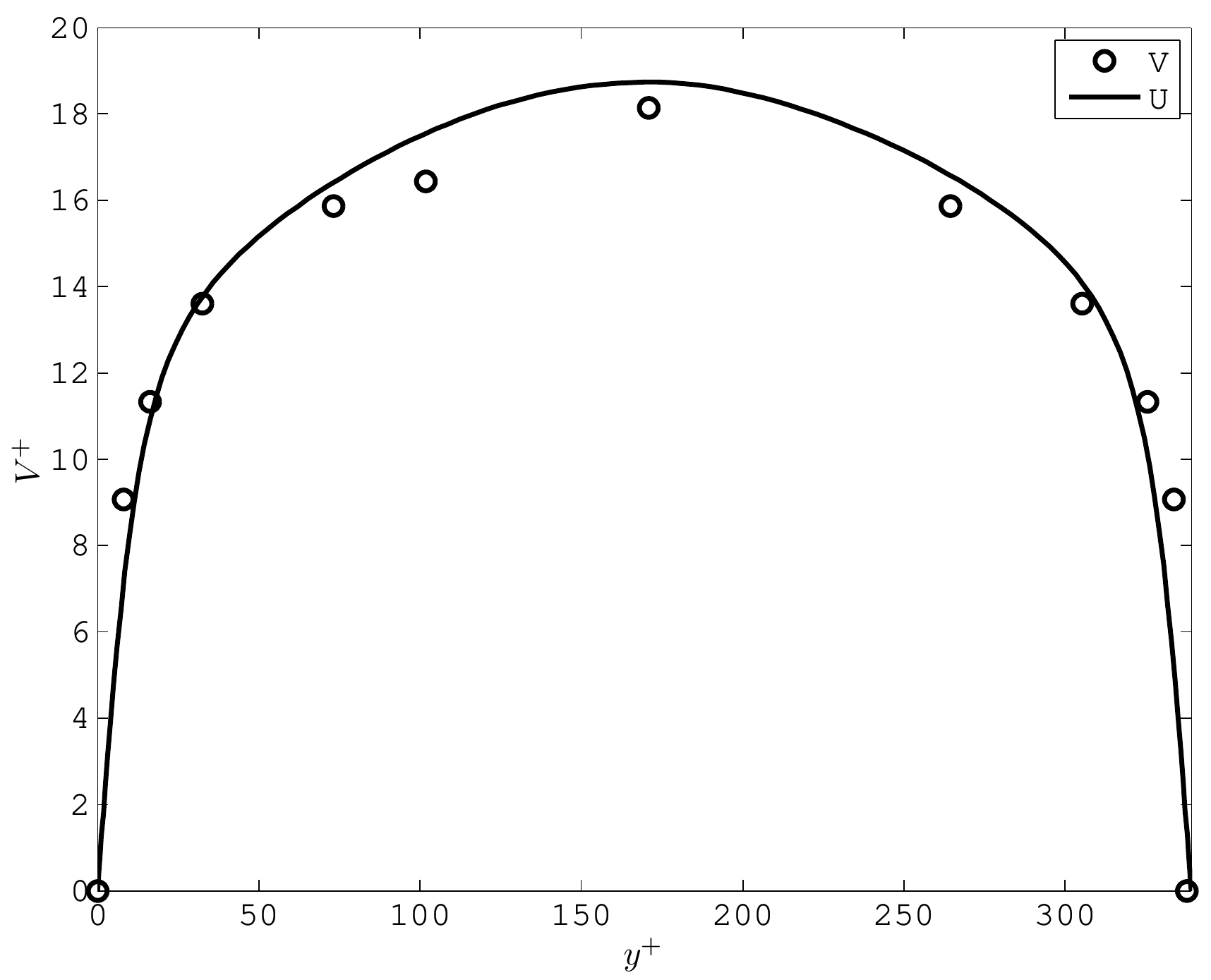}}
\caption{
Propagation velocity of FSLE field ($V^+$) and mean flow ($U^+$).
}
\label{Fig12}
\end{figure}
\end{center}

The profile of the propagation velocity is shown in 
figure \ref{Fig12}. The propagation velocity is 
very close to the mean flow velocity. 
The result shows that the maxima of the FSLE field,
that produce high values of $R_{uu}$ and where we expect to
find the ridges of the FSLE field, move with the flow. Hence,
one may conclude, as expected, that the FSLE ridges also move 
with the flow approximately as material
surfaces. 

\subsubsection{The 3d LCS.}
\label{3d_LCS_cf}

The previous description summarized the statistical properties
of the different structures appearing in an instantaneous FSLE
field. To make further progress we now extract
three-dimensional attracting LCSs in a region of the channel at
a series of time instants. The extraction domain had dimensions
$L_x^+ \times L_y^+ \times L_z^+ =103
\times  129 \times 124 $. The initial separation $d_0$ and
distance ratio $d_f/d_0$ were increased from the previous
calculation to improve the resolution and extract smoother
structures, but sacrificing a complete view of 3d LCS in the
turbulent channel. The extraction threshold was set to
$s=50000$, a compromise value between speed and cost of
extraction and continuity of the extracted surfaces. The FSLE
fields were calculated for an interval of $1.5$ time units with
a time step of $0.1$ units.

\begin{center}
\begin{figure}

\centerline{\includegraphics[width=0.9\textwidth, height=20cm]{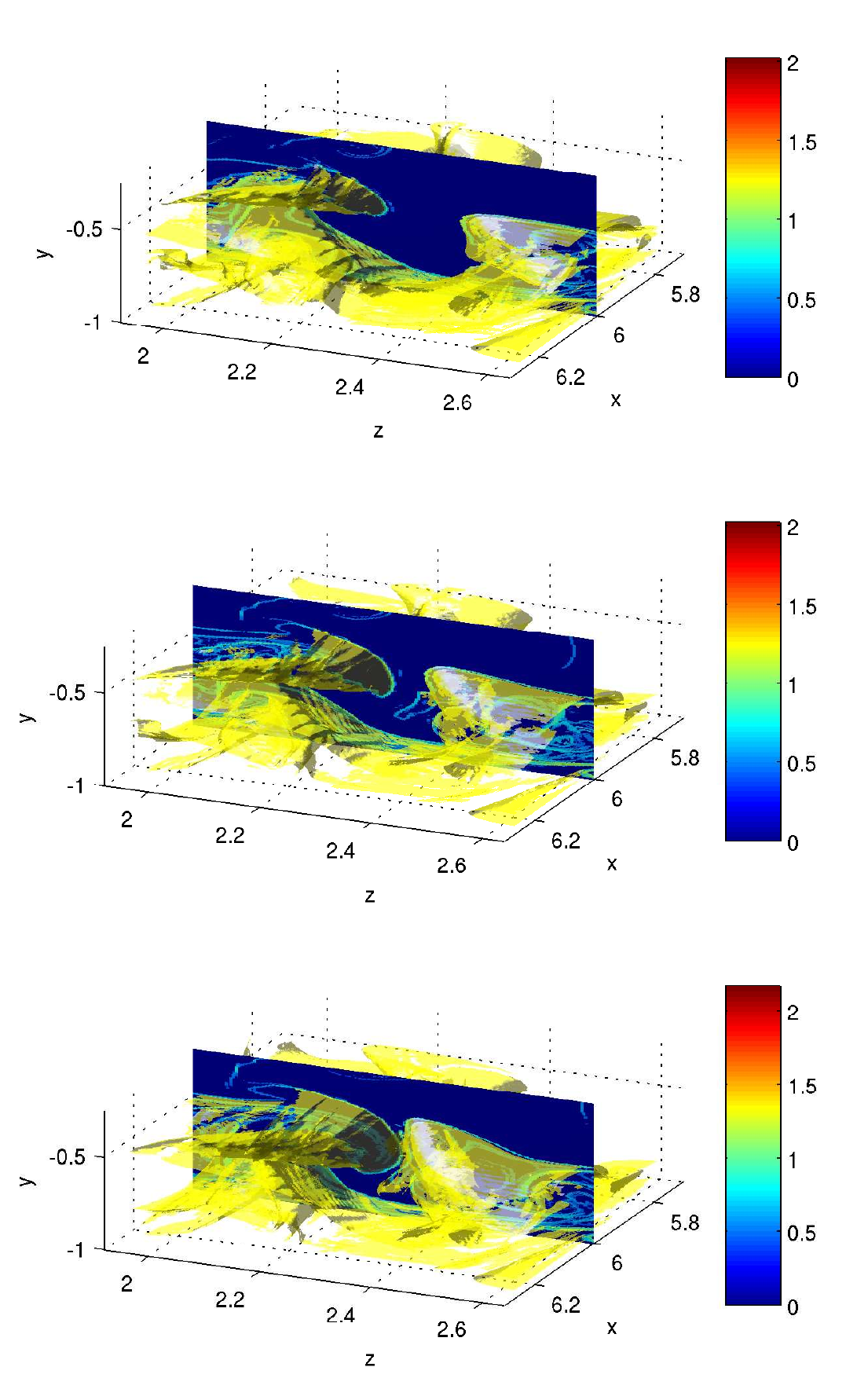}}
\caption{
3d attracting LCS in the channel flow together with a FSLE map at the fixed
plane $x=6.0$ ($x^+=1034$). Time goes from top to bottom, at intervals of 0.1 time units.
The flow direction is in the positive $x$ direction in each panel, and a wall is
at the bottom. The sequence
shows how one of the flow structures is advected and passes through the $x=6.0$ plane.
}
\label{Fig13}
\end{figure}
\end{center}

The 3d LCSs are rendered in figure \ref{Fig13}, in a sequence of time
instants, as they pass through the calculation domain. They
have a clearly 3d shape and move with the flow. The LCS seem to
create a boundary between the inner turbulent region and the
outer region that is practically devoid of FSLE. The highest
LCS have $\delta$-scale heights above the wall, and have a
distinct mushroom shape enclosing the regions of the channel
closer to the wall, where high FSLE values can be found. Near
the wall, the LCS adopt the shape of sheets parallel to it,
which reflects the high rates of shear that occur in that
region. These sheets form the base of the mushroom-shaped
excursions up to the channel center.

\section{Oceanic flow}
\label{sec:ocean_flow}

Contrarily to the turbulent flow of the previous section, large
scale oceanic flows, naturally turbulent, can be considered as
almost 2d due to rotation and stratification effects. This fact
makes the theory of 2d turbulence a very important tool to
understand the ocean processes that occur at large scales. The
main characteristic of 2d turbulence is the existence of an
inverse energy cascade, from the small to the large scales and
a direct enstrophy cascade. These cascades manifests themselves by
the creation of large coherent vortices, and by the process of
filamentation  by which strain regions in the boundaries of the
vortices produce lines of vorticity that are continuously
stretched and deformed by the flow, concentrating the vorticity
gradient in the small scales. This behavior is often observed
in oceanic flows thereby confirming the importance of the 2d
turbulent processes.

The results presented in this section were obtained in 
the Benguela ocean region, situated off the
west coast of southern Africa. It is characterized by a
substantial mesoscale activity in the form of eddies and
filaments, and also by the northward drift of Agulhas eddies.
The velocity data set comes from a regional ocean model
(ROMS) simulation of the Benguela Region \cite{LeVu2011}. 
Additional details on this work can be found in
\cite{Bettencourt2012}.

The three-dimensional FSLE fields were calculated for a $30$
day period starting September 17 of year 9, with snapshots
taken every $2$ days. The fields were calculated for an area of
the Benguela ocean region between latitudes 20\textdegree S and
30\textdegree S and longitudes 8\textdegree E to 16\textdegree
E. The calculation domain extended
vertically from $20$ up to $580$ m of depth. Both backward and
forward calculations were made in order to extract the
attracting and repelling LCS.

\begin{figure}
\centerline{\includegraphics[width=0.9\textwidth]{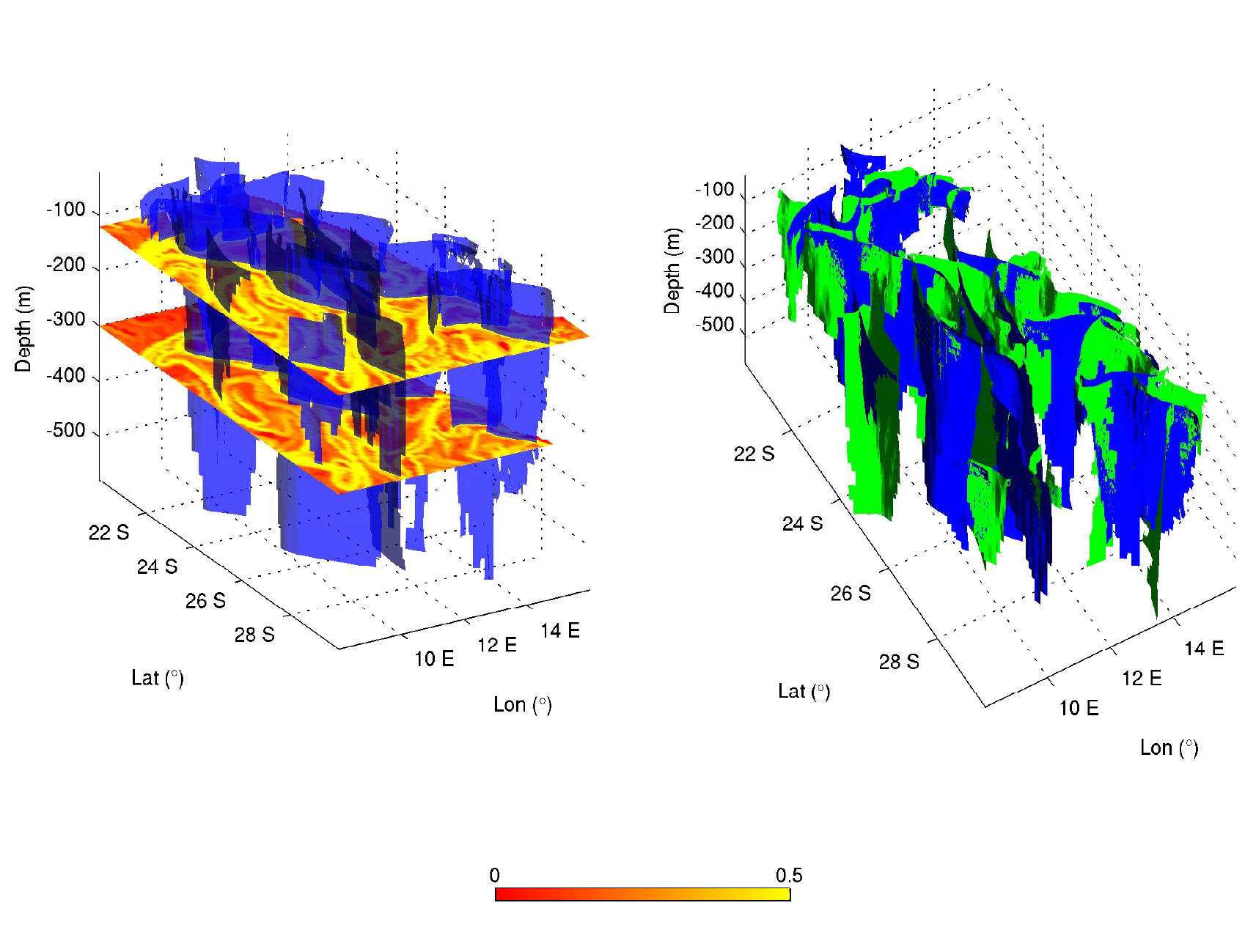}}
\caption{
3d LCS in the Benguela region for day 1 of the calculation period. Left panel
(from \cite{Bettencourt2012}):
Attracting LCS together with horizontal slices of the backward FSLE field at
120 m and 300 m depth. Right panel: Attracting (blue) and repelling (green) LCS.
Colorbar refers to colormap of horizontal slices in the left panel.
The units of the colorbar are $day^{-1}$.
}
\label{Fig14}
\end{figure}

In the left panel of figure \ref{Fig14} a snapshot of the
attracting LCSs for day 1 of the calculation period is shown.
The structures appear as thin vertical curtains, most of them
extending throughout the whole depth of the calculation domain.
The horizontal slices of the FSLE field in figure \ref{Fig14}
(left panel) show that the attracting LCS fall on the maximum
FSLE field lines, as in the case of the turbulent channel flow
(figure \ref{Fig13}). The FSLE fields themselves exhibit a variation in intensity
that decreases with depth, altough a local maximum is found
at $\sim 100$ m (not shown). The ridges also seem to be weaker
as the depth increases since for the same strength threshold, the
extracted portions of the ridges become less extent and eventually
vanish.
The atracting and repelling LCS (figure
\ref{Fig14}, right panel) populate the calculation region,
testifying the enhanced mixing activity that is known to occur
in that particular ocean region. The quite entangled ``web'' in
which attracting and repelling LCSs intersect mutually provides
the skeleton for the barriers and pathways controlling
transport \cite{dOvidio2004,Mancho2006b}.

\begin{figure}
\centerline{\includegraphics[scale=0.75]{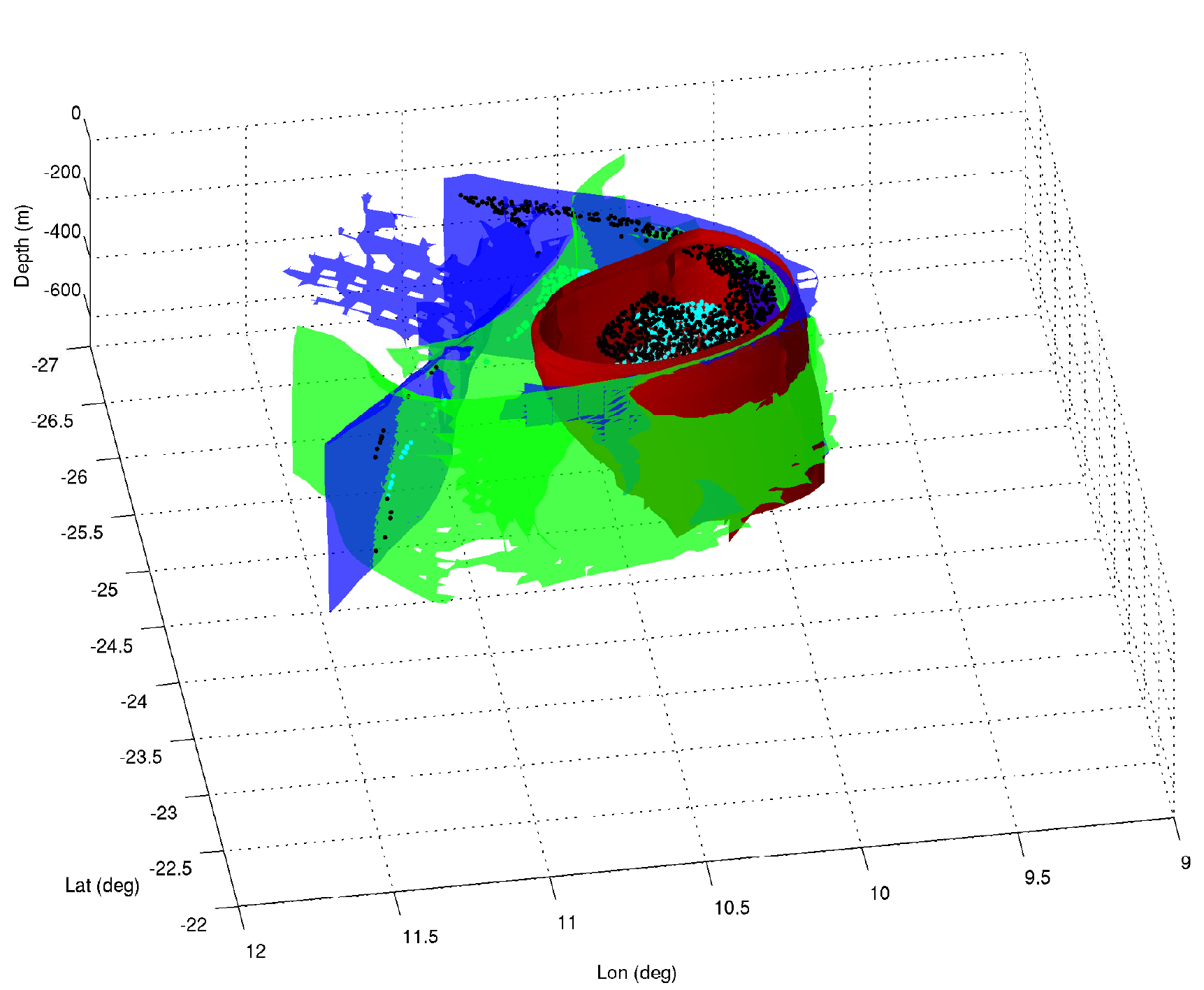}}
\caption{
Attracing (blue) and repelling (green) LCS on day 11 of the calculation period together
with Q-criterion isosurface at $Q=10^{-10}$ (red). The particles (black dots)
were released inside the eddy at day 1 at a depth of 50 m and are leaving now the eddy as a filament
along the upper part of the attracting LCS.
}
\label{Fig15}
\end{figure}

At this point, it may help to stress the
 differences between the Eulerian and Lagrangian detection of 
coherent structures.
This can be seen in figure \ref{Fig15} where the boundaries of a
mesoscale eddy are shown using the Q-criterion  and the
attracting and repelling LCS. The Q-criterion \cite{Hunt1988}
uses the second invariant of $\nabla\mathbf{u}$:
 \begin{equation}\label{Q1}
    Q=\frac{1}{2}(\|\mathbf{\Omega}\|^2-\|\mathbf{S}\|^2),
\end{equation}
where
$\|\mathbf{\Omega}\|^2=\textrm{tr}(\mathbf{\Omega}\mathbf{\Omega}^\mathrm{T})$,
$\|\mathbf{S}\|^2=\textrm{tr}(\mathbf{S}\mathbf{S}^\mathrm{T})$,
and $\mathbf{\Omega}$, $\mathbf{S}$ are the antisymmetric and
symmetric components of $\nabla\mathbf{u}$, to identify regions
where rotation dominates strain ($Q>0$), commonly identified
with coherent vortices, and strain dominated regions ($Q<0$).
We refer the reader to \cite{Jeong1995} and \cite{Haller2005}
for reviews and criticism of several Eulerian criteria.

Eulerian and Lagrangian measures limit approximately the same
region, but are substantially different. The Q-criterion is
related to the instantaneous configuration of the second
invariant of $\nabla\mathbf{u}$ and therefore conveys only
local information about fluid flow processes. The Lagrangian
perspective, on the other hand, provides an integration of the
temporal evolution of material properties of the flow, e.g.
material transport, and thus should give more meaningful
information about the processes that rely on ocean material
transport.

This issue can be further explored by looking at a
filamentation event (described more extensively in
\cite{Bettencourt2012}). A set of particles were released
inside the eddy at day 1 at a depth of 50 m. At day 11 of the
calculation period (see figure \ref{Fig15}), they have formed a
filament that is expelled from the eddy, so that particles
clearly cross the Q-criterion isosurface. This shows that the
Eulerian criteria is inadequate as an indicator of regions of
material transport in the flow. On the contrary, it can be
observed that the Lagrangian description of the eddy boundaries
does bear relation with material transport into and out of the
eddy, since the particle filament leaves the enclosed region
that we associate with the mesoscale eddy by following one of
the identified Lagrangian boundaries.

\section{Conclusions}
\label{sec:conclus}

Lyapunov exponents are useful to identify Lagrangian coherent
structures in turbulent flows. These constitute the pattern
determining the pathways of particle transport in the flow and
thus strongly influence the transport and mixing properties in
the fluid.

In this paper we have used a particular type of Lyapunov
exponents, the so-called Finite-Size Lyapunov exponents,  to
identify LCS in 3d flows. The finite size Lyapunov exponent was
used to measure the rate of streching of initially nearby fluid
particles in the flow domain and the Lagrangian coherent
structures where identified as the the ridges of the FSLE
field. These ridges were filtered in order to retain only the
strongest attracting or repelling structures.

In a  turbulent channel flow, the FSLE field is organized into
longitudinal structures close to the wall that develop into
sloping ones away from the wall. Correlations in the streamwise
and spanwise direction show the typical dimensions of these 
structures. They were found to be similar to the Eulerian coherent
structures that are known to exist in this same regions of the
turbulent channel. Specially, elongated streamwise vortices 
that move low speed fluid away from the wall into the channel core. 
In 3d, the LCSs appear as mushroom-shaped excursions of near-wall 
sheet-like structures of a scale comparable to the channel width. 
They separate the channel into an interior region, where the FSLE attains high
values, and an exterior region, showing low FSLE values. The
distribution of LCS in the turbulent channel resembles the
commonly accepted picture where upward excursions of near wall
fluid coexist with inward rushes of mid-channel irrotational
flow. Further work is necessary to elucidate the relations
between LCS and fluid transport in these type of flows, not
least because the visualization of 3d structures and transport
in turbulence is a complex and time-consuming subject.

In a quasi-2d mesoscale oceanic flow, the LCSs appear as
quasi-vertical surfaces highlighting the fact that dispersion
in this case is mainly horizontal. The high mixing activity can
be deduced from the proliferation of LCS in the flow domain and
their mutual intersection. These LCS were seen to provide
barriers and pathways to transport in the case of a mesoscale
eddy, contrary to Eulerian measures that failed to provide
indicative locations or directions of major transport events.

The main difference between these two 3d turbulent flows
with respect to the LCSs seems to be the fact that in the case of
oceanic flow, turbulence was limited to the horizontal plane wheras
in the channel flow case, turbulent fluctuations in all three space
directions had similar magnitude, thereby producing much more complex
3d shapes in this latter case. In the oceanic flow, vertical
motions have a tendency to be supressed by the combined effects of the
Earth's rotation and the stratification of the ocean. This results in
the aforementioned dominance of horizontal dispersion. The quasi-horizontal 
character of oceanic flows results in a phenomenology
of turbulence similar to that of 2d turbulence rather than to 3d turbulence. 

We note that there are fundamental differences between 
the Lagrangian and Eulerian
coherent structures, although they can actually have a common interpretation as
vortices or shear layers. Lagrangian
coherent structures have a clear impact in particle trajectories 
whereas Eulerian coherent structures are related to space/time coherency in, e.g.,
velocity signals and do not necessarily affect particles. 
In the above comparison, only the strongest FSLE features had a clear connection to
the features in the Eulerian distribution, which indicates that, inversely,
only the Eulerian features that live long enough or are strong enough
to affect particles in a discernible fashion will appear in the Lagrangian point
of view of coherent structures.  

The results shown in this paper highlight the usefulness of
Lyapunov analysis and dynamical systems theory as a tool to
study transport and mixing in fluid flows, through the concept
of Lagrangian coherent structures.

\section*{Acknowledgements}

This work was supported by Ministerio de Econom\'{\i}a y Competitividad
(Spain) and Fondo Europeo de Desarrollo Regional through
project FISICOS (FIS2007-60327). JHB acknowledges financial
support of the Portuguese FCT (Foundation for Science and
Technology) and Fundo Social Europeu (FSE/QREN/POPH) 
through the predoctoral grant SFRH/BD/63840/2009.

\section*{References}
\bibliographystyle{iopart-num}
\providecommand{\newblock}{}

\end{document}